\documentclass[a4paper,showpacs,nofootinbib]{revtex4-1}
\usepackage[utf8]{inputenc}
\usepackage{graphicx,bm,color,amssymb,amsmath}
\usepackage{slashed,verbatim}
\usepackage{subfigure}
\usepackage[colorlinks=true,linktocpage=true,linkcolor=blue,citecolor=blue]{hyperref}
\usepackage{placeins}
\usepackage{array}
\usepackage{xcolor} 
\usepackage[toc,page]{appendix}

\begin{document}
\hspace{15cm}TIFR/TH/21-4\\
\title{The role of $U(1)_A$ symmetry breaking in the QCD corrections to the pion mass difference}
\author{Mahammad Sabir Ali$^{1,2,a}$, Chowdhury Aminul Islam$^{3,b}$, Rishi Sharma$^{1,c}$}
\affiliation{$^1$ Department of Theoretical Physics, Tata Institute of Fundamental Research, Homi Bhabha Road, Mumbai 400005, India}
\affiliation{$^2$ School of Physical Sciences, National Institute of Science Education and Research, HBNI, Jatni, Khurda 752050, India}
\affiliation{$^3$ School of Nuclear Science and Technology, University of Chinese Academy of Sciences, Beijing 100049, China}
\email{$^a$ sabir@theory.tifr.res.in}
\email{$^b$ chowdhury.aminulislam@gmail.com}
\email{$^c$ rishi@theory.tifr.res.in}

\begin{abstract}
{The charged and neutral pion mass difference can be attributed to both the QED
and QCD contributions. The current quark mass difference ($\Delta m$) is the
source of the QCD contribution. Here, in a two flavour non-local NJL model, we
try to estimate the QCD contribution. Interestingly, we find that the
strength of the $U(1)_A$ symmetry-breaking parameter $c$ plays a crucial role
in obtaining the pion mass difference while intertwined with the current quark mass
difference. To obtain the QCD contribution for the pion mass difference, we scan the parameter space in $\{\Delta m,\; c\}$, and by comparing this with the
existing results, we constrained the parameter space. Further, using a fitted value of
$c$, we determine the allowed range for the $\Delta m$ in
the model. The model estimated $\Delta m$ ranges enable us to extract the chiral perturbation theory low-energy constant, $l_7$ and verify the dependence of the pion mass difference on $\Delta m$. We also find out its dependence on $c$ \textemdash\, it increases with the decreasing value of $c$, i.e., toward an axial anomaly restored phase.}
\end{abstract}

\maketitle

\section{Introduction} 
\label{sec:intro}

Symmetries of QCD play an important role in determining its low-energy
properties. In particular, the spontaneous breaking of the approximate chiral
symmetry from $SU(3)_L\times SU(3)_R$ to $SU(3)_V$ determines the low-energy spectrum of the theory and the interaction between the low-energy modes.
This is formalized in the low-energy effective field theory (EFT) well known as 
chiral perturbation theory ($\chi$PT) which describes the dynamics of the low-energy modes of QCD~\cite{Gasser:1984gg}.

An interesting role is played by the $U(1)_A$ transformations: equal chiral
transformations of the $u$, $d$, and $s$ quarks. This approximate classical
symmetry is broken~\cite{Adler:1969gk,Adler:1969er,Bell:1969ts} by quantum mechanical effects and thus is an anomalous symmetry. It leads to important effects like $\eta-\eta'$ mass splitting also known as the $U(1)$ problem of QCD. There are two major ways of dealing with it. One solution is obtained using instantons~\cite{tHooft:1976rip,tHooft:1986ooh}. The other is a topology-based solution obtained using a large $N_c$ limit without using the instantons~\cite{Witten:1979vv,Veneziano:1979ec,DiVecchia:1980yfw}. In effective models, one usually breaks this $U(1)_A$ using the `t Hooft determinant term~\cite{tHooft:1986ooh,Schafer:1996wv}.

On the other hand, for the two flavour QCD (considering only the
very light $u$ and $d$ quarks), $SU(2)_L\times SU(2)_R$
symmetry breaks to $SU(2)_V$ spontaneously and the low-energy theory is a chiral EFT
of $\pi$'s~\cite{Gasser:1983yg}. Effective models of QCD, for example the Nambu Jona-Lasinio
(NJL)~\cite{Nambu:1961tp,Nambu:1961fr} also capture aspects of QCD, in particular
its symmetries. In such models, $\pi$'s arise as collective modes of the theory. In this paper we will focus on two flavours.

The phenomenology of the NJL model and models inspired by NJL, in the iso-spin
symmetric limit with $m_u=m_d$, has been widely
studied~\cite{Klevansky:1992qe,Hatsuda:1994pi}. Many additional effects have
been studied in NJL-like models. For example, the effects of the Polyakov loop have
been included in the PNJL model~\cite{Fukushima:2003fw, Ratti:2005jh}. In
addition, situations of physical interest where external background fields
break iso-spin symmetry have been explored. As a first example, we point out
that the thermodynamics in the presence of a large magnetic field ($eB$) has been
analyzed (see Ref.~\cite{Shovkovy:2012zn} for a review). The difference in
the coupling of $u$ and $d$ quarks with the $B$ field breaks iso-spin. For
$eB\sim m_\pi^2$ the iso-spin symmetry-breaking effects on the $\pi$ masses is
comparable to their masses in the vacuum and therefore the effect on the low-energy
spectrum is substantial~\cite{Bali:2017ian}. As a second example, we point to studies of the
thermodynamics of QCD matter in the presence of a background iso-spin chemical
potential $\mu_I$~\cite{Buballa:2003qv}.

We know that the iso-spin symmetry in the QCD Lagrangian is also explicitly broken by the difference in the $u$, $d$
electrical charges (quantum electrodynamics (QED) effects) and due to non-zero
$\Delta m = m_d-m_u$ (sometimes called the ``mechanical contribution''~\cite{Dmitrasinovic:1995pr}). Both these
effects are small. The QED effects are small because of the small
$\alpha_{em}$. $\Delta m/\Lambda_{QCD}$ is comparably small. However, there are both effective model and lattice QCD studies exploring these effects. Some of the recent relevant NJL model calculations for meson masses can be found in Ref.~\cite{Osipov:2023rsg,Osipov:2023yif}. On the other hand, the
lattice calculations of the low-energy properties of QCD are becoming
increasingly accurate and multiple studies have calculated the low-energy
spectra including the iso-spin breaking effects~\cite{Brandt:2017oyy}. This behoves us to
revisit iso-spin breaking effects in NJL-like models.

One key observable is the difference between the charged and neutral pion mass.
In the presence of iso-spin symmetry, this mass difference is zero. At the
lowest order, one can consider the QED and the mechanical contributions
separately and add them. 

The contribution of the QED to the mass difference has been studied using
various approaches. The electromagnetic contribution to the pion mass
difference was first calculated using current algebra and QCD sum rules in
Ref.~\cite{Das:1967it} in the chiral limit. In Ref.~\cite{Gerstein:1967zza,Cook:1968ytb} the
authors calculated the finite quark mass corrections to the above calculation
for $m_u=m_d$.

In Ref.~\cite{Dmitrasinovic:1995pr} the value of the QED contribution and the
mechanical contribution have both been calculated in the two flavour NJL. The contribution coming from neutral pion mass correction ($O(\alpha m_\pi^2)$) plus the  mechanical contribution have been estimated  to be $1\%$.  In Ref.~\cite{Fujihara:2005wk} a similar calculation has been performed with the
additional analysis of the $\sigma$ mode and the finite temperature behaviour of the masses. 

More recently, iso-spin breaking due to QED as well as $\Delta m$ has been
studied on the lattice. The meson mass differences have been calculated in
lattice QCD (LQCD) simulations along with QED
interactions~\cite{Blum:2010ym,deDivitiis:2013xla,Basak:2013iw}. In these cases
disentangling the QED from the mechanical contributions is difficult and a
direct comparison with our results requires further assumptions.

From all these calculations we know that the QED contribution to the splitting is roughly $4.5$MeV and contributes most of the physical mass difference with the mechanical contribution $\lesssim 5\%$. Alternatively, if the QED contribution is known, the mechanical contribution can be obtained by subtracting the QED contribution from the pion mass difference from PDG~\cite{Tanabashi:2018oca}, 
\begin{eqnarray}
&&M_{\pi^0}=134.9770\pm 0.0005\ {\rm MeV},\nonumber\\
&&M_{\pi^+}-M_{\pi^0}=4.5936 \pm 0.0005\ {\rm MeV}.
\label{eq:pion_mass_exp}
\end{eqnarray} 

In this paper we will focus on the mechanical contribution. This has been calculated using $\chi$PT~\cite{Gasser:1984gg}. 
A more recent update of the results can be found in Ref.~\cite{Amoros:2001cp}. These results can be compared directly with the
results from our model and used to constrain its parameters.

For completeness we point to a few other works which use $\chi$PT to look at iso-spin
breaking effects on the meson spectra. Meson mass ratios and various decay
processes were used in Ref.~\cite{Donoghue:1993ha} to estimate the light quark
mass ratios. Ref.~\cite{Langacker:1974nm} used $\chi$PT with
two and three flavours to calculate the QED contributions to $\pi$ mass splitting and
$K$ mass splitting. The QED correction at finite temperature has been studied in 
Ref.~\cite{Nicola:2014eda}.

In all the analyses with the NJL model the four Fermi interaction was
considered to be of the standard form~\cite{Nambu:1961tp,Nambu:1961fr}
\begin{equation}
\Delta {\cal{L}} = \frac{G}{2}
\Bigl[(\bar{\psi}\psi)^2+(\bar{\psi}i\gamma^5\tau^a\psi)^2\Bigr]\;,
~\label{eq:standard}
\end{equation}
where $G$ is the coupling constant and $\tau$'s are the Pauli matrices.

Eq.~\ref{eq:standard} can be seen as a special case of a
$SU(2)_L\times SU(2)_R$ symmetric interaction term,
\begin{equation}
\begin{split}
\Delta {\cal{L}} &= \frac{G}{2}(1-c) 
\Bigl[(\bar{\psi}\psi)^2+(\bar{\psi}i\gamma^5\tau^a\psi)^2
+(\bar{\psi}i\gamma^5\psi)^2+(\bar{\psi}\tau^a\psi)^2\Bigr]\\
&+\frac{G}{2}c 
\Bigl[(\bar{\psi}\psi)^2+(\bar{\psi}i\gamma^5\tau^a\psi)^2
-(\bar{\psi}i\gamma^5\psi)^2-(\bar{\psi}\tau^a\psi)^2\Bigr],
\end{split}
~\label{eq:general}
\end{equation}
with $c=1/2$. The first piece in the interaction term is symmetric under $U(1)_A$ but the second piece breaks it which is known as the 't Hooft determinant term. In the absence of iso-spin  breaking, symmetry allows nonzero meanfield only to the $\bar{\psi}\psi$ channel, which is independent of the parameter $c$. Hence, the value of $c$ does not affect the free energy and any vacuum observables used to fit our model parameters. 

But it was pointed out in Ref.~\cite{Frank:2003ve} (also discussed in our
earlier paper~\cite{Ali:2020jsy}) that in the presence of iso-spin breaking effects (either
due to external backgrounds or due to $m_u\neq m_d$) the coefficient of the 't
Hooft determinant related to $c$ becomes important. In particular, the
gap equation for the difference of $\langle \bar{u} u\rangle$ and $\langle \bar{d} d\rangle$ is
sensitive to the value of $c$~\cite{Ali:2020jsy}.  

Consequently, the role of $c$ on the mass splitting between $\pi^{\pm}$ and $\pi^0$ should be taken into account when the mechanical contribution ($\Delta m$) is considered. This dependence has not been explored before our work. To estimate the pion mass splitting in the present scenario one needs to calculate the $\pi^\pm$ and $\pi^0$ masses within the regime of the non-local NJL model. The mass calculation for the charged pion is straightforward. On the other hand the calculation for $\pi^0$ is a bit subtle, as it mixes with the $\eta^*$ (isoscalar pseudoscalar)~\cite{Dmitrasinovic:1996fi}\footnote{One should note here that in two flavour this $\eta^*$ does not represent any physical particle.}. This mixing depends on both $c$ and $\Delta m$~\cite{Dmitrasinovic:1997te}. On calculating the masses we find that existing constraints on the mechanical contribution to the $\pi$ mass splitting, constrain the parameter space of $c$ and $\Delta m$ in the effective models. Furthermore, if we use the value of $c$ to match the lattice results~\cite{Bali:2012zg} for the splitting of the $u$ and $d$ chiral condensates in the presence of $eB$ (see Ref.~\cite{Ali:2020jsy} for details), we can constrain the value of $\Delta m$ using the splitting of pion masses. Using the constrained $\Delta m$ we also have an estimation of the range of the current quark mass ratio in the model. On comparing with the existing results~\cite{Amoros:2001cp,Donoghue:1993ha,Basak:2018yzz}, we find that our estimated ranges are reasonable.

One of the successful theories to describe pions is the chiral perturbation theory ($\chi PT$), which is constructed by expanding in the powers of external momenta and quark masses. This expansion contains unknown constants termed as the low-energy constants (LECs) that need to be fixed by matching observables to their physical/phenomenological allowed values. In a next-to-leading order two flavour $\chi PT$, there are a total of $7$ LECs ($l_{i,i=1,...,7}$)~\cite{Gasser:1983yg}; out of which, $l_7$ captures the isospin breaking due to nondegenerate current quark masses of $u$ and $d$ quarks. With the estimated range of $\Delta m$, we have calculated $l_7$ within our model and compared it with the available data from $\chi PT$ and LQCD calculations~\cite{Gasser:1983yg,GrillidiCortona:2015jxo,Boyle:2015exm,Frezzotti:2021ahg}. 

This LEC, $l_7$ relates the pion mass difference to $\Delta m$. $\chi PT$ predicts this dependence to be ${\cal O}(\Delta m)^2$~\cite{Gasser:1983yg}. We have verified this dependence in the present model (see also~\cite{Dmitrasinovic:1995pr}). Apart from that, we have also found a relation between $l_7$ and the 't Hooft determinant parameter, $c$. A decreasing value of $c$, which is equivalent to moving toward an axial anomaly restored phase, gives a higher value of $l_7$.

We use here the non-local NJL model, which has some advantages over its local
counterpart. Particularly, it can qualitatively mimic the running of the QCD
coupling constant and in turn regularizes the infinite integrals in the
theory~\cite{GomezDumm:2001fz}. As an outcome of this important feature, the
non-local NJL model has been quite successful while reproducing the inverse
magnetic catalysis (IMC) effect around the cross-over region~\cite{GomezDumm:2017iex}.

Our paper is organised as follows: First, we briefly describe the formalism
in section~\ref{sec:formalism} and then we reveal our findings in the result
section~\ref{sec:results} and finally in the section~\ref{sec:conclusion} we
conclude.

\section{Formalism}
\label{sec:formalism}
The most general two flavour NJL model with scalar interactions in the non-local
formalism in Euclidean space can be written as~\cite{Frank:2003ve,GomezDumm:2006vz,Hell:2008cc,Ali:2020jsy}
\begin{equation}
{\cal L}_{\rm NJL}=
{\bar{\psi}}\left(-i\slashed\partial+\hat{m}\right)\psi
-G_1\left\{j_a(x)j_a(x)+\tilde{j}_a(x)\tilde{j}_a(x)\right\}
-G_2\left\{j_a(x)j_a(x)-\tilde{j}_a(x)\tilde{j}_a(x)\right\}\;.
\end{equation}

The currents are given by 
\begin{equation}
j_{a}(x)/\tilde{j}_{a}(x)=\int d^4z\ {\cal H}(z) \bar{\psi}\left(x+\frac{z}{2}\right)\Gamma_{a}/\tilde{\Gamma}_{a}\,\psi\left(x-\frac{z}{2}\right),
\end{equation}
with $\Gamma=(\Gamma_0,\vec{\Gamma}_{j})=(\mathbb{I},i\gamma_5\vec{\tau}^j)$, $\tilde{\Gamma}=(\tilde{\Gamma}_0,\tilde{\vec{\Gamma}}_{{j}})=(i\gamma_5,\vec{\tau}^j)$, where $\vec{\tau}=(\tau^1,\tau^2,\tau^3)$ represents the Pauli matrices and ${\cal H}$(z) is the non-local form
factor in position space. The four components of $\Gamma$ ($\tilde{\Gamma}$) are labelled as $\Gamma_a$  ($\tilde{\Gamma}_a$). The combination of these currents with coupling
$G_1$ is axial $U(1)$ symmetric but the same with coupling $G_2$ is not. As
we are motivated to explore the effects of the current quark mass difference, $\hat{m}$ is
given by
\begin{equation}
\hat{m}=m_0\times\mathbb{I}-\frac{\Delta m}{2}\times \tau_3.
\end{equation} 
The non-zero current quark mass breaks the chiral symmetry and the remaining
$SU(2)_V$ is broken by the non-degenerate current quark masses to a $U(1)$
subgroup.

To integrate out the fermionic fields one can introduce auxiliary fields
associated with different currents. In the mean-field approximation,
symmetry considerations determine which of these auxiliary fields condense. In
an iso-spin symmetric scenario, the only allowed mean-field is associated with
the current for $\Gamma_0$, which then couples to the fermions with strength $G_1+G_2$. Whereas in an iso-spin symmetry broken scenario the auxiliary field associated with $\tilde{\vec{\Gamma}}_{j}$ can acquire non-zero mean-field values,
coupled to fermions with coupling proportional to $G_1-G_2$. Here, we take $G_1+G_2=G/2$ and $G_1-G_2=\Delta G/2=(1-2c)G/2$.  

The bosonized action is given by
\begin{equation}
\Omega=\int d^4x\left\{ \frac{S_a(x)S_a(x)}{2G}+\frac{\tilde{S}_a(x)\tilde{S}_a(x)}{2\Delta G} \right\}-\ln \det (A),
\label{eq:action}
\end{equation}
where the $S(x)/\tilde{S}(x)$ are the auxiliary fields associated with the currents $j(x)/\tilde{j}(x)$ as follows
\begin{equation}
    S_a(x)=-(G_1+G_2)j_a(x)\hspace{0.8cm} {\rm and} \hspace{0.8cm} \tilde{S}_a(x)=-(G_1-G_2)\tilde{j}_a(x).
    \label{eq:aux_field}
\end{equation}
The fermion inverse propagator in momentum space is given by
\begin{equation}
A(p,p')=\left(-\slashed p +\hat{m}\right)(2\pi)^4\delta^4(p-p')+g\left(\frac{p+p'}{2}\right)\left\{\Gamma_aS_a(p-p')+\tilde{\Gamma}_a\tilde{S}_a(p-p')\right\},
\end{equation}
where $g(q)$ is the non-locality form factor in momentum space, obtained by Fourier transforming ${\cal H}(x-y)$ as follows
\begin{equation}
    g(q)=\int d^4z\, e^{iq.z}{\cal H}(z).
\end{equation}

Lorentz symmetry ensures that $g$ can only be a function of Lorentz invariant quantities constructed from $q$, i.e., $q^2$. Here, we choose this form factor to be Gaussian, $g(q)=e^{-q^2/\Lambda^2}$~\cite{Ali:2020jsy}. Denoting the mean field value of the $\Gamma_0$ bilinear operator as $S_0(x)=\sigma$ and the value of the $\tilde{\Gamma}_3$ operator as $\tilde{S}_3(x)=\Delta\sigma$, respectively, one can minimize the action by simultaneously solving the two gap equations associated with these mean-fields which are given below:

\begin{equation}
\frac{\sigma}{G}-2N\sum_{f}\int\frac{d^4p}{(2\pi)^4}\,g(p)\,\frac{M_f(p)}{p^2+M_f^2(p)}=0\,\,{\mathrm{and}}
\label{gap1}
\end{equation}
\begin{equation}
\frac{\Delta\sigma}{(1-2c)G}-2N\sum_{f}(-1)^{f+1}\int\frac{d^4p}{(2\pi)^4}\,g(p)\,\frac{M_f(p)}{p^2+M_f^2(p)}=0,
\label{gap2}
\end{equation}

where $f$ is the flavour index and takes values $1\,(u\,{\mathrm{quark}})$ and $2\,(d\,{\mathrm{quark}})$, and $2N=4N_c$ where $N_c$ is the number of colour. The constituent quark masses ($M_f$) are given by
\begin{equation}
M_f(p)=m_0+\left(f-\frac{3}{2}\right)\Delta m+g(p)\left(\sigma+(-1)^{f+1}\Delta\sigma\right).
\label{constituent_mass}
\end{equation}

Using the solutions of the two gap equations, the condensates of individual
flavours can be obtained as
\begin{equation}
\langle{\bar\psi}_f\psi_f\rangle=-2N\int\frac{d^4p}{(2\pi)^4}
\left[\frac{M_f(p)}{p^2+M_f^2(p)}-\frac{m_f}{p^2+m_f^2}\right].\label{eq:cond}
\end{equation}
It is to be noted here that the mass dimension of the quantity $\langle{\bar\psi}_f\psi_f\rangle$ is three, and its calculated value is negative. We will be mainly interested in calculating the average and difference of condensates which are given as
\begin{align}
\Sigma_{\rm Ave}=\left({|\langle\bar{u}u\rangle|}^{1/3}+{|\langle\bar{d}d\rangle|}^{1/3}\right)/2,\hspace{1cm}{\rm and}\hspace{1cm} \Sigma_{\rm Diff}={|\langle\bar{u}u\rangle|}^{1/3}-{|\langle\bar{d}d\rangle|}^{1/3},
    \label{eq:cond_ave_dif}
\end{align}
where $\langle\bar{u}u\rangle$ and $\langle\bar{d}d\rangle$ are the $u$ and $d$ quark condensates, respectively. This sets up the basic formalism used to calculate the condensate average and
difference shown in Sec.~\ref{sec:results}.  

\subsection{Mesonic propagator}
To make the action quadratic in quark fields, we perform the Hubbard Stratanovich~\cite{1957SPhD.2.416S,PhysRevLett.3.77} transformation and introduce auxiliary fields given in Eq.~\ref{eq:aux_field}. To study the mesonic properties (for example, their masses) within this
formalism we need to expand the above action (Eq.~\ref{eq:action}) in powers of
mesonic fluctuation to obtain the propagators. The mesonic fluctuations are identified as the fluctuations of the associated auxiliary fields around their meanfields. Hence, we expand the auxiliary fields in Eq.~\ref{eq:aux_field} around their meanfield values to obtain the coefficient of the quadratic term, which is equivalent to the inverse propagator. The relevant auxiliary fields to calculate the pion and $\eta^*$ mesons are
\begin{eqnarray}
S_{i\gamma_5}(x)&=&\delta\eta^*(x)\,,
\end{eqnarray} 
\begin{eqnarray}
S_{i\gamma_5{\tau^3}}(x)&=&\delta\pi^0(x)\;\;\;{\mathrm{and}}\;\;\;
S_{i\gamma_5\tau^{1,2}}(x)=\delta\pi^{1,2}(x)\,.
\end{eqnarray} 
Note that we have separated the $\tau^3$ component of the mesons from the $\tau^1$,
$\tau^2$ components because this component is split for $m_u\neq m_d$.  In this paper we
will focus on the difference between the $\pi^{0}$ and the $\pi^{\pm}$ (or
equivalently the $\pi^{1,2}$) masses. 

The interpretation of the meson fields is immediate. The pseudo-scalar
iso-vector mesons ($\pi$'s) are the lightest degrees of freedom (they are massless
in the chiral limit). The scalar mesons are heavy. The case of the $\eta^*$ meson
is interesting. For $c=0$, the Lagrangian is $U_A(1)$ symmetric and the $\eta^*$
is degenerate with the $\pi^0$. $U_A(1)$ breaking lifts the mass of the $\eta^*$~\cite{Dmitrasinovic:1996fi}.
In the two flavor analysis, the $\eta^*$ cannot be directly identified with mesons in 
nature. It can be seen as a fictitious meson representing an admixture of the 
$\eta$ meson and the $\eta'$ in three flavor QCD. This is why we will not
analyze it further but as we shall see below, the $\eta^*$ mode will play a role
in determining the mass of the $\pi^0$. 
The inverse mesonic propagator can be obtained by integrating out the fermion
fields.

For the $\pi^{\pm}$ fields, one finds~\cite{Klevansky:1992qe,GomezDumm:2006vz} that the inverse 
propagator for the charged pions ($\pi^+$ or $\pi^-$) in momentum space is proportional to
\begin{eqnarray}
{\cal{G}}^{\mathrm{ch}}(p^2) &=& \frac{1}{G} - \, \int \frac{d^4 q}{(2 \pi)^4}\
g^2(q) {\rm{tr}}\left\{
i\gamma^5 \tau^{1,2}
\frac{1}{\slashed{q}+\frac{1}{2}\slashed{p}+[M]}
i\gamma^5 \tau^{1,2}
\frac{1}{\slashed{q}-\frac{1}{2}\slashed{p}+[M]}\right\}\\
    &=&\frac{1}{G} - \, 4 \,N_c\int \frac{d^4 q}{(2 \pi)^4}\
g^2(q) \left\{\frac{ \left[ (q^+ \cdot q^-) + M_u(q^+)
	M_d(q^-)\right]}{\left[ (q^+)^2 + M_u^2(q^+) \right]
	\left[ (q^-)^2 + M_d^2(q^-)\right]}+(u\leftrightarrow d)\right\}\;.
\label{eq:pion_prop_c}
\end{eqnarray}
Here we have defined $q^\pm = q \pm p/2$. $[M(p)]$ is a $2\times 2$ matrix of
the form ${\rm{diag}}(M_u(p),M_d(p))$.  The form factor $g$ appears because of
the definition of the non-local currents~\cite{GomezDumm:2006vz}.

The calculation for the $\pi^0$ inverse propagator is more
subtle because the $\pi^0$ can mix with $\eta^*$ for $m_u\neq m_d$~\cite{Dmitrasinovic:1996fi}. The inverse
propagator for the two fields in momentum space has the formal structure
\begin{equation}
[{\cal{G}}^0(p^2)] = [G]^{-1} - [\Pi(p^{2})]\;.
\label{eq:pi0_eta_propagator}
\end{equation}

One should note here the difference between ${\cal{G}}^{\mathrm{ch}}(p^2)$ and $[{\cal{G}}^0(p^2)]$, the first one represents the inverse propagator for the charged pions whereas the last one is the matrix containing the inverse propagators for $\pi^0$, $\eta^*$ and their mixing terms. Also, 
\begin{equation}
[G] = 
\left[
\begin{array}{cc}
G & 0\\
0 & \Delta G
\end{array}
\right]\label{eq:g_matrix}
\end{equation}
and
\begin{equation}
[\Pi(p^2)] = 
\left[
\begin{array}{cc}
\Pi^{\pi^0\pi^0}(p^2) & \Pi^{\pi^0\eta^*}(p^2)\\
\Pi^{\pi^0\eta^*}(p^2) & \Pi^{\eta^*\eta^*}(p^2)
\end{array}
\right]\;.
\end{equation}

Here,
\begin{eqnarray}
{\Pi^{\pi^0\pi^0}}(p^2) = {\Pi^{\eta^*\eta^*}}(p^2) =
\, 4 \,N_c\sum_{f} \int \frac{d^4 q}{(2 \pi)^4}\
g^2(q) \frac{  \left[ (q^+ \cdot q^-) + M_f(q^+)
	M_f(q^-)\right]}{\left[ (q^+)^2 + M_f^2(q^+) \right]
	\left[ (q^-)^2 + M_f^2(q^-)\right]}\,\, {\rm and}
\label{eq:pion_prop_n}
\end{eqnarray}
 
\begin{eqnarray}
{\Pi^{\pi^0\eta^*}}(p^2) = {\Pi^{\eta^*\pi^0}}(p^2) = \, 4 \,N_c\sum_{f} (-1)^{f+1} \int \frac{d^4 q}{(2 \pi)^4}\
g^2(q) \frac{  \left[ (q^+ \cdot q^-) + M_f(q^+)
	M_f(q^-)\right]}{\left[ (q^+)^2 + M_f^2(q^+) \right]
	\left[ (q^-)^2 + M_f^2(q^-)\right]}.
\label{eq:pion_n_prop}
\end{eqnarray}

Finally, equating the determinant of $[{\cal{G}}^0(p^2)]$ (given in Eq.~\ref{eq:pi0_eta_propagator}) to zero, we obtain two pole masses \textemdash$\;$ one is for the $\pi^0$ (the lighter mode) and the other for the $\eta^*$ (the heavier mode) which now contains the mixing effects.

\subsection{Parameters} 
In Ref.~\cite{Ali:2020jsy}, we have fitted all the parameters along with the $U(1)_A$ symmetry-breaking strength to LQCD data. The results given below are obtained using the LH parameter set (Table I in Ref.~\cite{Ali:2020jsy}), for which the phase diagram in $T-eB$ has better agreement with LQCD. At physical pion mass $m_\pi=135$ MeV, the parameters of the model are fitted to $\Sigma_{\rm{Ave}}=221.1$ MeV in the iso-spin symmetric limit, and pion decay constant $f_\pi=92.9$ MeV. The fitted parameters are, current quark mass $m_0=6.94$ MeV, $G=29.38/\Lambda^2$ and $\Lambda=605.05$ MeV. We have obtained $c_{\rm fit}$ for which the condensate difference agrees well with LQCD result, given as\footnote{Please check Eq. (54) in Ref.~\cite{Ali:2020jsy} and the discussion therein.}
\begin{align}
c_{\rm fit}=0.149^{+0.103}_{-0.029}.
\label{eq:cfit}
\end{align}
We will be varying $\Delta m$ as a parameter in our model and we have checked that it $(\Delta m)$ has no significant effect on these zero temperature observables $(m_\pi,\,\Sigma_{\rm{Ave}}\, \mathrm{and}\, f_\pi)$. On the other hand clearly, the observables $|\langle\bar{u}u\rangle|^{1/3}-|\langle\bar{d}d\rangle|^{1/3}$ and the pion mass difference are significantly affected by both $c$ and $\Delta m$.

\section{Results}
\label{sec:results}
We now explore the effects of the current quark mass difference on the pion mass difference intertwined with the axial symmetry-breaking effects. The iso-spin symmetry is explicitly broken by $\Delta m$. We will be particularly interested in the QCD contribution or the mechanical contribution (i.e., the effect of $\Delta m$) in the pion mass difference.

We will first explore the sensitivity of the $\pi$ mass difference on $c$ in general. The choice of the values of $c$ is motivated as follows. $c=1/2$ corresponds to the standard NJL model.  $c=0$ is a natural lower limit because $c<0$ has unphysical
consequences~\cite{Ali:2020jsy}. We focus on the region between $c=0$ and $c=1/2$ because
this is favored by results in the $eB$ dependence of the $u$, $d$ condensate
difference~\cite{Ali:2020jsy}. Note that $c=0$ corresponds to $U(1)_A$
symmetric interactions.

\begin{figure}[!htbp]
	\includegraphics[scale=0.72]{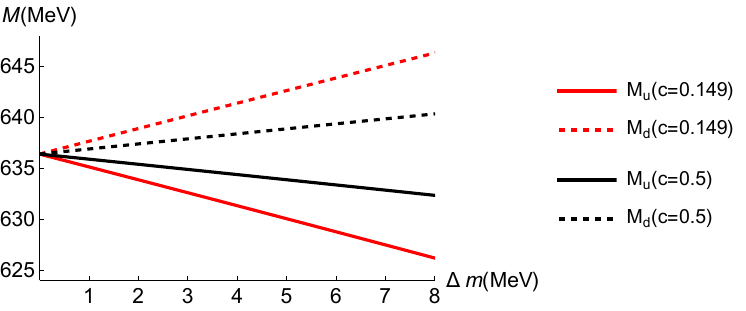}\hspace{0.6cm}
	\includegraphics[scale=0.66]{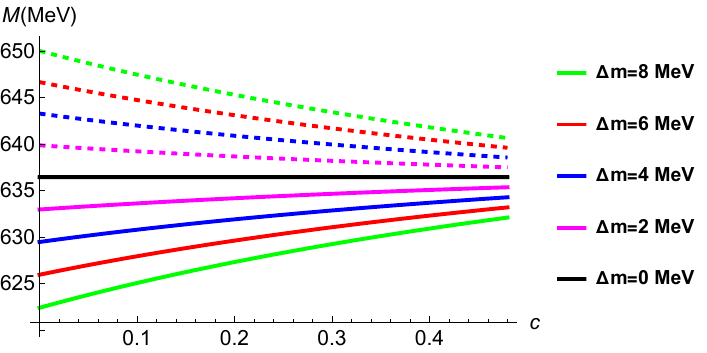}
    \caption{The constituent masses for $u$ (solid lines) and $d$ (dashed
    lines) quarks: In the left panel for two different values of $c$ as
    function of mass difference ($\Delta m$) and in the right panel for
    different values of $\Delta m$ as a function of $c$.}
	\label{fig:mass_dm_c}
\end{figure}

First, we use the gap equations, Eqs.~\ref{gap1} and~\ref{gap2}, to find the mean fields $\sigma$ and $\Delta\sigma$. Then we use Eq.~\ref{constituent_mass} to find the $p$ dependent constituent mass. The left panel of Fig.~\ref{fig:mass_dm_c} presents the $\Delta m$ dependence of constituent masses of $u$ and $d$ quarks (at zero momentum ($M_f(0)$))\footnote{The value of the constituent mass obtained here is relatively large as compared to the local version~\cite{Klevansky:1992qe,Hatsuda:1994pi}. One should note that this is for zero momentum and is consistent with other studies~\cite{Eichmann:2015kfa}. The effective mass runs with momentum as opposed to the constant effective mass in local cases. It should also be noticed that the condensate's value, which is of more physical relevance, is within the range of phenomenologically acceptable values in our calculation.}. We observe that the difference between the $u$ and $d$ quarks constituent masses increases with the increase of $\Delta m$ and the difference is higher for smaller values of $c$. For $c=0.5$ the difference between the constituent masses is exactly equal to the values of $\Delta m$, which can be understood from the Eqs.~\ref{gap2}
and~\ref{constituent_mass}. These two equations also explain the higher
constituent mass difference for a given value of $\Delta m$ for $c$ other than
$0.5$, provided the mean-field $\Delta\sigma$ is negative which is the case
here.

In the right panel of Fig.~\ref{fig:mass_dm_c}, we have plotted constituent masses as a function of
$c$ for different values of $\Delta m$. This is a different point of view of the left panel of the same figure, and it can be seen that the constituent masses for a nonzero value of $\Delta m$ depend nonlinearly on $c$, at least for the smaller values. The black solid line represents the $c$-independent equal mass of $u$ and $d$ quarks for zero current quark mass difference. This corroborates our statement in the introduction that in the absence of iso-spin breaking (here, $\Delta m=0$) the value of c does not affect the observables. With the constituent masses in our hands we are ready to compute the observables like the chiral condensate and the $\pi$ masses.

\begin{figure}[!htbp]
	\includegraphics[scale=0.7]{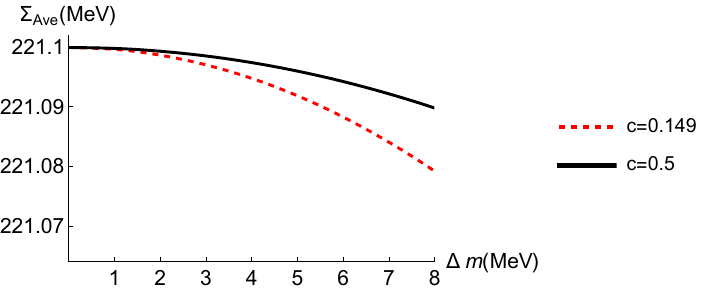}
	\includegraphics[scale=0.7]{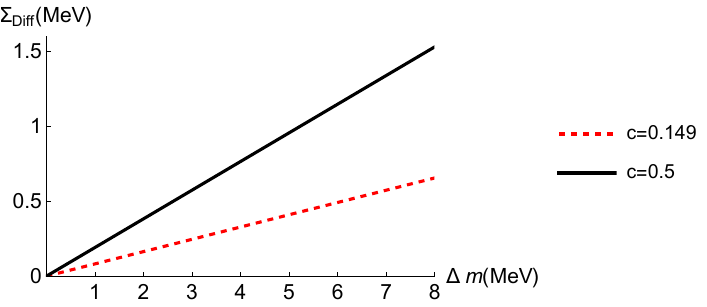}
	\caption{The condensate average (left) and difference (right) for two different values of $c$ as a function of mass difference ($\Delta m$).}
	\label{fig:cacd_dm}
\end{figure}

Using the $u$, $d$ constituent masses we found, in Fig.~\ref{fig:cacd_dm}
we present the condensate average and difference for a couple of representative values of $c$ as a function of mass difference ($\Delta m$). 

In the left panel, we can see that the condensate average is not
very sensitive to $\Delta m$, though smaller $c$ values increase the effect of
$\Delta m$. On the other hand, the figure in the right panel shows the condensate difference as a function of $\Delta m$ for two different values of $c$. The condensate difference increases with the increase of $\Delta
m$ for a fixed value of $c$, and it also increases as $c$ increases with $\Delta m$ kept fixed. One might expect that the condensate will be proportional to the constituent mass, which implies that the $d$ condensate will be greater than the $u$ condensate. But due to the regularization scheme, we have obtained the opposite behavior in condensate difference. A decrease in $c$ increases the gap between the constituent masses. But as the regularization is unaffected by $c$, and $d$ quark gets higher subtraction (for larger current mass, see Eq.~\ref{eq:cond}), giving rise to the decreases in the condensate difference as we decrease $c$.

\begin{figure}[!htbp]
	\includegraphics[scale=0.66]{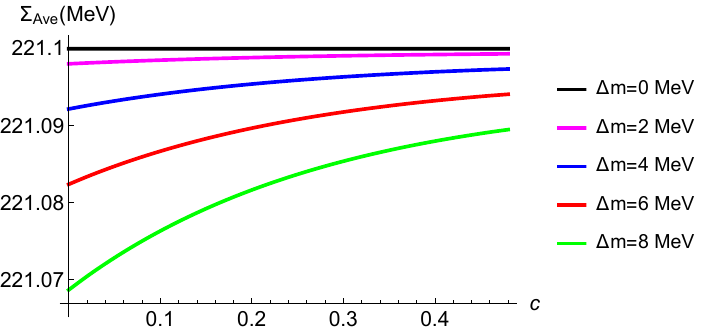}
	\includegraphics[scale=0.66]{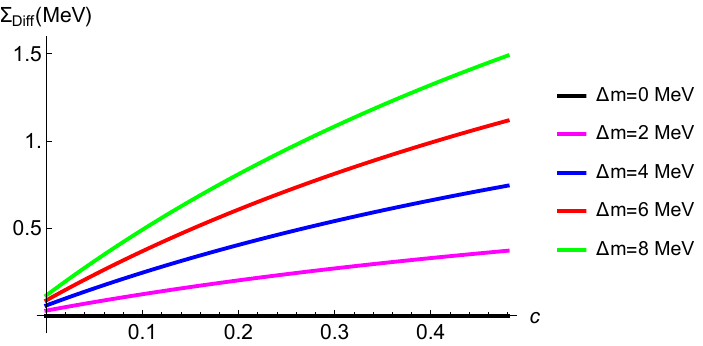}
	\caption{The condensate average (left) and difference (right) for different values of $\Delta m$ as a function of $c$.}
	\label{fig:cacd_c}
\end{figure}

In Fig.~\ref{fig:cacd_c}, we have plotted the condensate (both average and the
difference) as a function of $c$ for different values of $\Delta m$. The left
panel of this figure correlates with the left panel of Fig.~\ref{fig:cacd_dm}.
In the right panel, on the other hand, the plot of condensate difference as a
function of $c$ reflects some interesting observations. The condensate
difference decreases with decreasing $c$ for a given value of $\Delta m$. To put things into perspective, we would like to mention here the effect of
another iso-spin symmetry-breaking agent, external magnetic field ($eB$), on
the condensate difference. There we observed that it increases with increasing
$eB$ and/or decreasing $c$~\cite{Ali:2020jsy}. In the present study, the decreasing $c$ has the opposite effect as for increasing $\Delta m$ on the condensate difference.

\begin{figure}
	\includegraphics[scale=0.7]{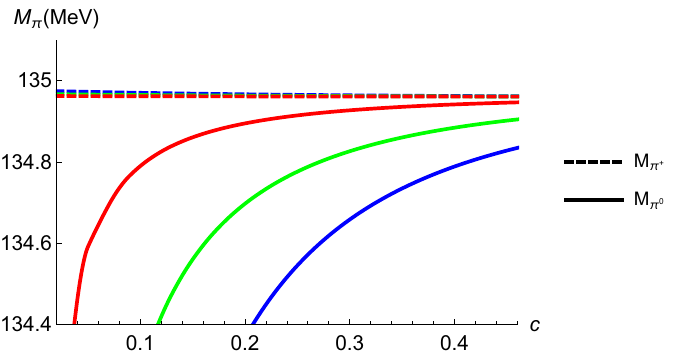}
    \caption{The pion masses (without the QED effects) as a function of $c$ for different values of
    current quark mass difference ($\Delta m$). The dashed and solid lines represent
    the charged ($\pi^+$) and neutral pions ($\pi^0$) masses, respectively. Red, green and blue data correspond to $\Delta m=2,\, 4\,\, {\rm and}\ 6\, {\rm MeV}$, respectively.}
	\label{fig:pi_mass_diff}
\end{figure}

The pion masses as a function of $c$ for fixed $\Delta m$ are presented in
Fig~\ref{fig:pi_mass_diff}. In our formalism, we found out that the charged pion
mass is almost independent of $\Delta m$ and $c$. On the other hand, the neutral
pion mass decreases as one increases $\Delta m$ and/or decreases $c$. The
enhanced decrease in the neutral pion mass for small $c$ is due to mixing with
$\eta^*$. At $c=0$ the $\eta^*$ becomes degenerate with the $\pi^0$ meson
giving rise to maximal mixing effect and as one increases $c$ the mixing effect
decreases. As the $\eta^*$ meson in two flavour does not correspond to any
physical particle we have not shown the $\eta^*$ mass. It is to be noted that
the QED correction has not been taken into account for the charged pion mass
value, which is supposed to lift its value to the physical one and gives a
relatively smaller correction to the neutral pion
mass~(Eqs.~\ref{eq:pion_prop_n} and~\ref{eq:pion_prop_c}).
  
\begin{figure}
\begin{center}
	\includegraphics[scale=0.6]{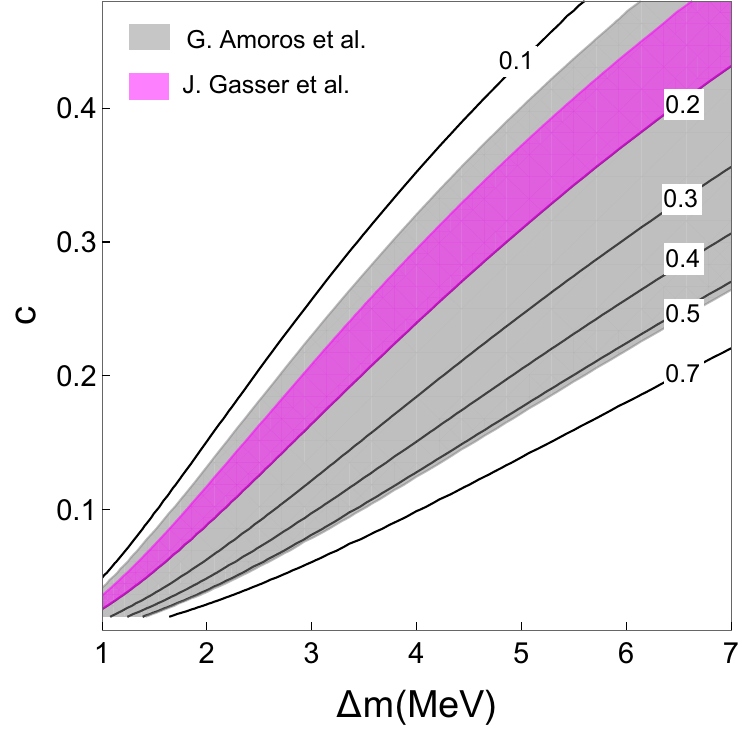}
	\caption{Contour plot of pion mass difference in $\Delta m$ and $c$. The magenta and gray bands are from $\chi$PT calculation~\cite{Gasser:1984gg} and~\cite{Amoros:2001cp}, respectively.}
	\label{fig:contour}
\end{center}	
\end{figure}

In Fig.~\ref{fig:contour}, we have presented a contour plot depicting the
different lines of constant $\Delta M_\pi$ and their corresponding ranges in
$\Delta m$ and $c$. To constrain the parameter space in $\{\Delta m,\; c\}$ in the
model, we further compare our findings with the existing results in the
literature. We focus on results for the mechanical contribution to the $\pi$
mass splitting calculated via $\chi$PT and from lattice QCD.

Authors of Ref.~\cite{Gasser:1984gg} have calculated the QCD contribution in
pion mass difference using chiral perturbation theory. They obtained
$(M_{\pi^+}-M_{\pi^0})_{\rm QCD}=0.17(03)$ MeV. The $\chi$PT calculation was
revisited in Ref.~\cite{Amoros:2001cp} and they obtained it to be $0.32(0.20)$ MeV due to the current quark mass difference\footnote{As we observe the mechanical contribution to the pion mass difference in Ref.~\cite{Amoros:2001cp} is different from that in Ref.~\cite{Gasser:1984gg}. There are two sources for this difference: one is due to the inclusion of the mass-correction in the Kaon electromagnetic mass difference and the other comes from the consideration of $\mathcal{O}(p^6)$ effects~\cite{Amoros:2001cp}.}. These results from
$\chi$PT are presented in the same plot using magenta~\cite{Gasser:1984gg} and gray~\cite{Amoros:2001cp} bands in
Fig.~\ref{fig:contour}.

\begin{figure}
	\includegraphics[scale=0.7]{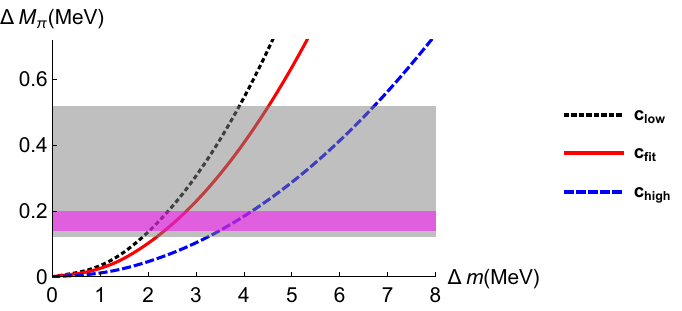}
	\includegraphics[scale=0.6]{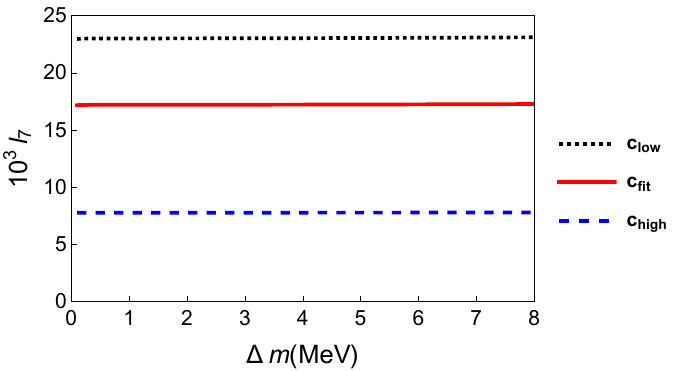}
	\caption{Left panel: the pion mass difference as a function of $\Delta m$ for the fitted $c$ value from Ref.~\cite{Ali:2020jsy}. The magenta and gray bands are from $\chi$PT calculation~\cite{Gasser:1984gg} and~\cite{Amoros:2001cp}, respectively. Right panel: the $\Delta m$ dependence of $l_7$. For details, please check the text.}
	\label{fig:Mass_diff_cfit}
\end{figure}

In our previous work~\cite{Ali:2020jsy}, we have constrained the parameter $c$
using some lattice QCD (LQCD) data. There, we have fitted $c$ to match the
condensate difference in the model with that of LQCD~\cite{Bali:2012zg} in presence of $eB$
at $T=0$. We have used the model parameters to reproduce the
JLQCD~\cite{Fukaya:2007pn} observables for which the fitting of $c$ is
reasonably good~\cite{Ali:2020jsy}. Using that fitted $c$ value
(given in Eq.~\ref{eq:cfit})\footnote{There have been other
efforts~\cite{Frank:2003ve,Boomsma:2009yk} to constrain the parameter $c$,
which are similar to the value we obtained.}, we have shown the pion mass
difference as a function of $\Delta m$ in Fig.~\ref{fig:Mass_diff_cfit}. In the same figure we have shown the
estimations of $\Delta M_\pi$ from two different $\chi$PT
calculations~\cite{Gasser:1984gg,Amoros:2001cp} which give us the opportunity
to extract an allowed range for $\Delta m$. 
We take $\Delta M_\pi=0.17(03)$ MeV from Ref.~\cite{Gasser:1984gg} and $\Delta M_\pi=0.32(20)$ MeV from Ref.~\cite{Amoros:2001cp}. Then the allowed ranges for $\Delta m$ are obtained as 
\begin{align}
\begin{split}
\Delta m &=2.59\left(_{-0.24}^{+0.22}\right)\left(_{-0.35}^{+1.26}\right) {\rm{MeV\; for}} \text{~\cite{Gasser:1984gg}} {\rm{,\;and}}\\
\Delta m &=3.55\left(_{-1.37}^{+0.97}\right)\left(_{-0.48}^{+1.73}\right) {\rm{MeV\; for}} \text{~\cite{Amoros:2001cp}}\;.
\end{split}
\end{align} 
In both cases the central values of $\Delta m$ are obtained from the central value of the $\Delta M_\pi$ at the best-fit $c$ value (Eq.~\ref{eq:cfit}). The first error is due to the spread in $\Delta M_\pi$~\cite{Gasser:1984gg,Amoros:2001cp} while $c$ is kept fixed at the best-fit value, and the second error is due to the spread in $c$ (Eq.~\ref{eq:cfit}) while $\Delta M_\pi$ is kept fixed at the central values.

In view of the recent progress on the calculation of iso-spin violating effects
in lattice QCD it is worth making a comment on the comparison of our results
with these. Ref.~\cite{Blum:2010ym} has obtained the pion mass difference to be
$(M_{\pi^+}-M_{\pi^0})_{\rm QED}=4.50(23)$ MeV. In this calculation~\cite{Blum:2010ym}, the
charged pion mass is calculated with non-degenerate $(\Delta m\ne0)$ light quark masses but
for the neutral pion, they have considered degenerate quark masses $(\Delta m=0)$. On the other hand,
in our study, we have considered the effect of non-degenerate quark mass 
for both the charged and neutral pions. Thus, because of 
their (\cite{Blum:2010ym}) consideration of degenerate current quarks for neutral mesons,
we cannot make a direct comparison with their calculation.  Nevertheless, to get a feel for the numbers
this value can be subtracted from the experimental value
(Eq.~\ref{eq:pion_mass_exp}) to obtain an estimate of the QCD correction, which
is $0.09(0.23)$ MeV. With the further assumption that in their calculation the
$\Delta m$ has a negligible effect on the neutral pion mass, a comparison with
our result is possible. One should also note here that with the consideration
of the error, the range for the QCD contribution extends up to negative values.

Ref.~\cite{deDivitiis:2013xla} is a more recent LQCD calculation with iso-spin
symmetry broken by quark mass difference and their respective electric charges.
They obtained $(M^2_{\pi^+}-M^2_{\pi^0})_{\rm QED}=1.44(13)(16)\times10^3$
MeV$^2$. Using the experimental values for the sum of the pion
masses (Eq.~\ref{eq:pion_mass_exp}) one can obtain the QED correction for
$(M_{\pi^+}-M_{\pi^0})_{\rm QED}=5.24(0.75)$ MeV. With this number at hand we
can estimate the QCD contribution in the same way as mentioned above, which is
found to be $-0.65(0.75)$ MeV. In all these studies, overall the error bar for
QED correction is still substantial and that leads to a wide range for the QCD
correction once the QED values are subtracted from the experimental ones. We will see in our model that the LEC $l_7$ is positive, and this implies that the mechanical contribution should be positive. This is also consistent with other $\chi$PT and LQCD calculations.

\subsection{$l_7$} 

We know that in $\chi$PT, the iso-spin breaking is characterized by the low-energy constant (LEC) $l_7$~\cite{Gasser:1983yg}. The estimate of plausible ranges of $\Delta m$ in our calculation provides us with the scope of calculating the same in the present model as well. The spread in the current quark mass difference also allows us to look for $\Delta m$ dependence of $l_7$, if any. To evaluate $l_7$ we use the pion mass splitting induced by the square of the current quark mass difference ($(\Delta m)^2$). The relation is given by
\begin{align}
M_{\pi^+}^2-M_{\pi^0}^2=(m_u-m_d)^2\frac{2B_0^2}{f_\pi^2}l_7,
\label{eq:l7_eq}
\end{align} 
where $B_0$ can be written in terms of $M_\pi$ as $M_\pi^2\simeq(m_u+m_d)B_0$. Using the above relation, we can estimate the $l_7$ as 
\begin{align}
l_7=17.2^{+5.8}_{-9.4}\times 10^{-3}.
\end{align}
The range given in $l_7$ in the above equation is due to the spread in $c$ \textemdash\, the upper spread comes from the value $c_{\rm low}$ and the lower spread from the value $c_{\rm high}$ (see Fig.~\ref{fig:Mass_diff_cfit}). $\chi$PT has estimated the $l_7$ to be $5\times 10^{-3}$~\cite{Gasser:1983yg}, whereas Ref.~\cite{GrillidiCortona:2015jxo} has estimated it to be $7(4)\times 10^{-3}$. One of the recent calculations, which fitted pseudoscalar meson mass and decay constant to partially quenched $\chi$PT to next-to-leading and next-to-next-to-leading order~\cite{Boyle:2015exm}, has estimated it to be $6.5(3.8)\times 10^{-3}$. Another recent LQCD calculation for $2+1+1$ dynamical quark flavour obtains it to be $2.5(1.4)\times 10^{-3}$~\cite{Frezzotti:2021ahg}. Thus we can safely remark that our estimated value using a simple 2-flavour NJL model is in the ballpark.

Other than estimating the value of $l_7$, there are some other important observations that we would like to emphasise here. From the $\chi$PT calculation, we know that the nonzero contribution to $\Delta M_\pi$ from $\Delta m$ is ${\cal O}(\Delta m^2)$ (Eq.~\ref{eq:l7_eq}). To verify this dependence we plotted $l_7$ as a function of $\Delta m$ for the fitted $c$ value, as shown in the right panel of Fig.~\ref{fig:Mass_diff_cfit}. It is evident that $l_7$ is independent of $\Delta m$ for a fixed value of $c$. Thus, we stress here that albeit the obtained $l_7$ value in the model is somewhat larger as compared to the $\chi$PT or LQCD, it is worth noting that the dependence predicted through $\chi$PT can be successfully verified. Another important point to note is the dependence of such LEC like $l_7$ on the axial anomaly. Its value increases as we move toward restoring the anomaly, i.e., decreasing value of $c$, as also evident from the right panel of Fig.~\ref{fig:Mass_diff_cfit}.

\section{Summary and Conclusion}
\label{sec:conclusion}
It is known that the charged and neutral pion masses are different. The
difference arises from both QED and QCD contributions. The QED contribution
emerges out of the different charges of $u$ and $d$ quarks. On the other hand,
the QCD contribution comes into play because of the difference in current
masses for $u$ and $d$ quarks ($\Delta m$), which is also known as the
mechanical contribution. Because of different quark compositions for
charged and neutral pions, these QED and QCD effects contribute and make the masses
of charged and neutral pions as different as we observe.

Both these effects have been calculated using different effective QCD models,
EFTs, as well as LQCD calculations. Some of the works have estimated the contribution
from one type of effect and the other contribution has been calculated by using
the experimental values. We here explored particularly the QCD contribution
i.e., the effect of current quark mass difference using a 2-flavour non-local
NJL model in the presence of axial symmetry [$U(1)_A$] breaking parameter $c$. This
handed us the unique and interesting opportunity of investigating the effect of
current quark mass difference on the pion mass difference intertwined with the
axial symmetry-breaking effects. The consideration of the axial
$U(1)_A$ breaking term becomes particularly relevant in the presence of some iso-spin
breaking agents like magnetic field and/or iso-spin chemical potential, as we have shown in our last work~\cite{Ali:2020jsy}.

Our main findings suggest that the charged pion mass is almost insensitive to the iso-spin symmetry-breaking strength, whereas the neutral pion mass decreases with increasing $\Delta m$. The parameter $c$ does not have any non-trivial impact on the charged pion mass, whereas, for the neutral pion, it has some effect. The effect is significant considering the fact that the effect of $\Delta m$ (the QCD contribution) on the pion mass difference is small as compared to the total pion mass difference. To make our calculation consistent, we have taken $\pi-\eta^*$ mixing into account. As we can see from Eq.~\ref{eq:g_matrix}, the $\eta^*$ mass explicitly depends on the parameter $c$ through the $\Delta G$ term. While we explore the whole range of physically meaningful $c$ values, we specifically give attention to the value that we have obtained by fitting some lattice QCD observables in our previous
work~\cite{Ali:2020jsy}.

We have estimated the pion mass difference for a wide range of $\Delta m$ and
physically meaningful $c$ values and presented that in a contour plot along
with the existing results, mainly from chiral perturbation theory.  We have
obtained an allowed zone in the parameter space in $\{\Delta m,\,c\}$. To
constrain the parameter space further we have used a fitted $c$ value and
obtained only a $\Delta m$ dependent pion mass difference. The allowed ranges
for $\Delta m$ are found to be $2.59\left(_{-0.24}^{+0.22}\right)\left(_{-0.35}^{+1.26}\right)$ MeV~\cite{Gasser:1984gg} and $3.55\left(_{-1.37}^{+0.97}\right)\left(_{-0.48}^{+1.73}\right)$ MeV~\cite{Amoros:2001cp} for the two references, respectively.

From the obtained range of $\Delta m$ we can also calculate the range for
the current quark mass ratio in our model. The above-mentioned ranges of $\Delta m$ can be translated into the quark mass ratios, and we obtain $m_u/m_d=0.69\left(^{+0.02}_{-0.02}\right)\left(^{+0.04}_{-0.15}\right)$ and $0.59\left(^{+0.14}_{-0.08}\right)\left(^{+0.04}_{-0.14}\right)$, respectively. On comparing our
estimations with the existing results, we find that they are a bit higher than
that found in some of the older calculations ($0.30$)~\cite{Donoghue:1993ha}
and are within the range provided by a more recent estimate
$0.46(0.09)$~\cite{Amoros:2001cp}. One might look into
Ref.~\cite{Basak:2018yzz}, particularly in Fig. 19, for different existing
estimations of $m_u/m_d$. Our estimated ranges are comparable to the values
quoted there.

To explore further, we have calculated $l_7$, a low-energy constant (LEC) in $\chi PT$, in our model. It essentially captures the information about the effect of $\Delta m$ on $\Delta M_\pi$. Our observation is as follows: for a fixed parameter set, $l_7$ is very sensitive to the parameter $c$ and is independent of $\Delta m$. This confirms that $\Delta M_\pi$ depends on $\Delta m$ quadratically. Its nontrivial dependency on $c$ implies that the breaking of axial symmetry could play a very important role in physical observables like pion mass difference. We have obtained $l_7$ to be $17.2^{+5.8}_{-9.4}\times 10^{-3}$, which is slightly higher than the existing results~\cite{Gasser:1983yg,GrillidiCortona:2015jxo,Boyle:2015exm,Frezzotti:2021ahg} but falls within their spread.
\\

{\bf Acknowledgement}: The authors would like to thank Veljko Dmitrašinović and Hiranmaya Mishra for their critical comments and suggestions regarding the manuscript. M.S.A. and R.S. would also like to acknowledge many useful discussions with Sourendu Gupta and Subrata Pal. M.S.A. would like to acknowledge the support provided by TIFR, Mumbai, where most of the work was done, and would also like to acknowledge the support from NISER for the current position of a Senior Project Associate. M.S.A and R.S. would also like to acknowledge the support of the Department of Atomic Energy,
Government of India, under Project Identification No. RTI 4002. C.A.I. would like to acknowledge the financial support by the Chinese Academy of Sciences President's International Fellowship Initiative under Grant No. 2020PM0064 and partial help from the Fundamental Research Funds for the Central Universities, China.

\bibliography{ref}

\end{document}